\documentstyle[prb,floats,aps]{revtex}
\begin{document}
\title{Exact Solution of Heisenberg-liquid models with long-range coupling}
\author{Yupeng Wang$^{1,2,3}$ and P. Schlottmann$^2$}
\address{$^1$Institute of Physics \& Center for Condensed Matter Physics, Chinese Academy of Sciences, 
Beijing 100080, People's Republic of China}
\address{$^2$Department of Physics, Florida State University, 
Tallahassee, FL 32306, USA}
\address{$^3$Center for Advanced Study, Tsinghua University, Beijing 100084, People's Republic of China}

\maketitle
\date{\today}
\maketitle
\begin{abstract}
We present the exact solution of two Heisenberg-liquid models of 
particles with arbitrary spin $S$ interacting via a hyperbolic 
long-range potential. In one model the spin-spin coupling has the 
simple antiferromagnetic Heisenberg exchange form, while for the 
other model the interaction is of the ferromagnetic Babujian-Takhatajan 
type. It is found that the Bethe ansatz equations of these models 
have a similar structure to that of the Babujian-Takhatajan spin 
chain. We also conjecture the integrability of a third new spin-lattice 
model with long-range interaction.
\end{abstract}

\section{introduction}

Spin fluctuations are a common feature of strongly correlated electron 
systems such as high $T_c$ superconductors and heavy fermion compounds. 
Exchange interactions of the Heisenberg type, originally introduced 
to describe the properties of insulating magnets, have now been 
realized playing a central role in doped Mott insulators and heavy 
fermion systems. Two related models are the $t-J$ model and the Kondo 
lattice model. In these systems, the spin exchange among the itinerant 
electrons or between the conduction electrons and the local moments 
induce fascinating physical phenomena, such as very large mass 
enhancements and non-Fermi liquid behavior. For some low-carrier-density 
systems, especially in low dimensions, the small number of carriers 
and their reduced mobility do not provide an effective mechanism for 
screening and long-range interactions should be considered. A special 
class of integrable systems with long-range interactions are the 
Calogero-Sutherland model (CSM)\cite{calo,suth} and its $SU(N)$ 
generalizations.\cite{poly1,kawa,ha} The lattice versions of these 
models, which generically are referred to as Haldane-Shastry models, 
\cite{hald} can be obtained by freezing the orbital dynamics within 
a proper scheme.\cite{poly2} In this letter, we present two integrable 
models of particles with arbitrary spin interacting via a 
Calogero-Sutherland potential with a hyperbolic space dependence. 
In case I the spin-spin coupling is of the simple Heisenberg exchange 
form, i.e. ${\vec S}_i\cdot{\vec S}_j$, while in case II, the 
interaction is of the Babujian-Takhatajan model (BTM) type.\cite{babu} 
Finally, based on the structure of the Bethe ansatz equations (BAE), 
an $SU(2)$-invariant high-spin chain model, which corresponds to the 
long-range-coupling generalization of the BTM, \cite{babu} is 
conjectured to be integrable. 

The structure of the present paper is the following: In the subsequent 
section, we construct the model Hamiltonians. The energy spectra and 
the BAE are derived via the so-called asymptotic Bethe ansatz 
(ABA).\cite{suth} In sect.III, we study the thermodynamics as well as 
the ground state properties.  Concluding remarks are given in sect.IV.

\section {Models and Bethe ansatz}

The general form of the model Hamiltonian we shall consider is
\begin{eqnarray}
H = -\sum_{j=1}^N \frac{\partial^2}{\partial x_j^2} + 2\sum_{i<j}^N
\frac{\gamma^2 g_{ij}}{\sinh^2\gamma(x_i-x_j)} , \label{H}
\end{eqnarray}
where $x_j$ is the position of the particle $j$, $\gamma>0$ is a 
real constant, and $g_{ij}$ is a spin-dependent coupling, which 
for our two cases is: 
\begin{eqnarray}
g_{ij} = 2{\vec S}_i\cdot{\vec S}_j + 2S(S+1),
\end{eqnarray}
for model $I$ and 
\begin{eqnarray}
g_{ij} = -4SQ_{2S}({\vec S}_i\cdot{\vec S}_j) \equiv 
[2S - {\hat l}_{ij}] [2S + 1 - {\hat l}_{ij}],
\end{eqnarray}
for model $II$, where 
\[ {\hat l}_{ij} = \sqrt{\frac14 + ({\vec S}_i+{\vec S}_j)^2} -
\frac12 , \] 
which has eigenvalues $0, \cdots, 2S$ and $Q_{2S}$ is just the 
Babujian polynomial.\cite{babu} Hence, in case $I$ the spin coupling 
is a simple antiferromagnetic exchange, while in case $II$ it is 
ferromagnetic and of the Babujian-Takhatajan form. It is interesting 
to compare model $I$ to the $t-J$ model. Their physics is quite 
similar, although in the present model the coordinate $x_j$ is 
continuous while it is restricted to a lattice in the $t-J$ model, 
and that here the interaction is of variable range. For some high 
$T_c$ compounds, the next-nearest-neighbor spin exchange is believed 
to be relevant. In the present model the range of the interaction is 
governed by the free parameter $\gamma^{-1}$. For instance, for 
sufficiently large $\gamma$ the interaction is short-ranged, while the 
long-range Calogero-Sutherland interaction is readily recovered in the 
limit $\gamma\to 0$. We note that in both cases, $I$ and $II$, the 
total spin is conserved in the two-body scattering processes and the 
Hamiltonian (\ref{H}) has a basic $SU(2)$-invariance.

The asymptotic Bethe ansatz (ABA)\cite{suth} is a powerful tool to 
derive the energy spectrum for models with nonlocal interactions, like 
e.g. the CSM. If the model is integrable it is sufficient to know the 
asymptotic long-distance behavior, i.e.~the scattering phase shifts, 
which can be obtained without the full knowledge of the wave functions. 
The ABA also assumes that the particles are asymptotically free, i.e., 
that they do not form charge bound states. The ABA only needs the 
two-body phase shifts (or two-body scattering matrix), but requires 
an independent proof of the integrability of the model Hamiltonian. 
The integrability of the present model is related to the unitary 
matrix model of Minahan and Polychronakos,\cite{poly3} who have 
shown that the CSM with a coupling strength $g_{ij}=J_{ij}J_{ji}$ 
is exactly soluble. In Ref. [\onlinecite{poly3}] $J_{ij}$ was chosen 
to be the $SU(q)$ angular momentum. For the $2S$-fold $SU(2)$ case, 
we can map $J_{ij}J_{ji}$ to $({\vec S}_i+{\vec S}_j)^2$, which is 
of more interest to condensed matter problems and corresponds to our 
case $I$. In fact, a potential ${\vec J}{}^2/r^2$ can be generated 
naturally from the dynamics of two free particles in three dimensions 
in spherical coordinates. Such a potential describes the angular 
dynamics of the center of mass of two {\it free} particles. If the 
angular momentum is considered a ``spin'', the problem is reduced 
to a one-dimensional one and the scattering matrix must be 
factorizable, which already infers the integrability of many 
particles in one dimension with such a potential. Hence, for case 
$I$ our choice of $g_{ij}$ is quite natural. On the other hand, 
case $II$ can be realized by a proper quantization of $J_{ij}$ using 
projection operators.\cite{poly3}

To derive the two-body $S$-matrix, we note that the two-particle 
scattering occurs in $2S+1$ different channels of total spin $l$ 
ranging from $0$ to $2S$. In a given channel of spin $l$, the 
coupling strength is $l(l+1)$ for case $I$ and $(2S-l)(2S-l+1)$ for 
case $II$. For convenience we will express our results in terms of 
the operator ${\hat l}_{ij}$ defined above. Below we study the two 
cases separately.

\subsection{Model $I$}

The two-body transmission $S$-matrix can be derived following 
Sutherland's method.\cite{suth} Consider the two-particle case. 
The asymptotic wave function for $|x_1-x_2|\to\infty$ can be 
written as
\begin{eqnarray}
\Psi(x_1,x_2)\to A_{12}(12)e^{ik_1x_1+ik_2x_2}+A_{12}(21)
e^{ik_2x_1+ik_1x_2},
\end{eqnarray}
for $x_1<x_2$ and
\begin{eqnarray}
\Psi(x_1,x_2)\to A_{21}(21)e^{ik_1x_1+ik_2x_2}+A_{21}(12)
e^{ik_2x_1+ik_1x_2},
\end{eqnarray}
for $x_1>x_2$, where $k_1, k_2$ are the asymptotic momenta carried 
by the particles and the $A's$ are constant coefficients, which 
determine the $S$-matrices. For example, $S_{1,2}(k_1-k_2)=A_{21}(21)
/A_{12}(12)$. By solving the Schr\"odinger equation for two 
particles we obtain
\begin{eqnarray}
S_{i,j}(k_i-k_j)=-\frac{\Gamma(1+ik_{ij})}{\Gamma(1-ik_{ij})}
\frac{\Gamma(1+{\hat l}_{ij}-ik_{ij})}{\Gamma(1+{\hat l}_{ij}+i
k_{ij})}P_{ij},
\end{eqnarray}
where $k_{ij}=(k_i-k_j)/2\gamma$, $\Gamma(x)$ is the gamma function 
and $P_{ij}$ is the spin permutation operator. In fact, $S_{i,j}$ is 
a polynomial of ${\vec S}_i\cdot{\vec S}_j$ of order $2S$ and is 
proportional to the Lax-operator of the BTM.\cite{babu} Hence, it 
satisfies the Yang-Baxter relation\cite{yang} $S_{i,j}S_{i,m}S_{j,m}
=S_{j,m}S_{i,m}S_{i,j}$. For example, when $S=1/2$, we have
\begin{eqnarray}
S_{i,j}=\frac{k_{ij}+i(1/2+2{\vec S}_i \cdot {\vec S}_j)}{k_{ij}-i},
\end{eqnarray}
while for $S=1$ it is
\begin{eqnarray}
S_{i,j}=
-\frac{2{\vec S}_i\cdot{\vec S}_j(1-ik_{ij}+{\vec S}_i\cdot{\vec S}_j)
-k_{ij}^2+ik_{ij}-2}{(ik_{ij}+1)(ik_{ij}+2)} .
\end{eqnarray}
Because the system is non-diffractive, the wave function must 
asymptotically be given by plane waves $\Psi(x|Q) \sim \sum_P 
\Psi(P|Q) \exp[i\sum_{j=1}^N x_{Q_j}k_{P_j}]$ when $|x_i-x_j|
\to\infty$ for all pairs $i,j$, where $x_{Q_1}<\cdots<x_{Q_N}$, 
and $P$ and $Q$ are permutations of $(1,2,\cdots, N)$. With the 
periodic boundary conditions of the wave function, the momenta $k_j$ 
 are determined by a set of Yang eigenvalue equations\cite{yang}
\begin{eqnarray}
S_{j,j-1} \cdots S_{j,1} S_{j,N} \cdots S_{j,j+1} \Psi(P|Q) 
= e^{-ik_jL} \Psi(P|Q).
\end{eqnarray}

For convenience, we introduce an auxiliary spin ${\vec S}_0$ and 
define $S_{0,j}=S_{0,j}(\lambda-k_j)$, so that $S_{0,j}(0)=
P_{0j}$. Further we define the $(2S+1)\times(2S+1)$ monodromy matrix
\begin{eqnarray}
T_0(\lambda)=S_{0,j}\cdots S_{0,N}S_{0,1}\cdots S_{0,j+1}.
\end{eqnarray}
By summing over the auxiliary spin of the matrix $T_0(\lambda)$, 
it is easy to show that the trace 
$$tr_0T_0(k_j)=S_{j,j-1}\cdots S_{j,1}S_{j,N}\cdots S_{j,j+1}$$
is just the transfer matrix of the BTM\cite{babu} with an extra 
multiplying factor and lattice disorder (we consider a continuum 
model). The eigenvalue problem Eq.(9) can be solved following the 
standard method of the algebraic Bethe ansatz,\cite{babu,fadd} 
\begin{eqnarray}
&&e^{ik_jL} = \left\{ \prod_{n=1}^{2S} \prod_{i,i\neq j}^N 
\frac{k_j-k_i-2in\gamma}{k_j-k_i+2in\gamma} \right\}
\prod_{\alpha=1}^M \frac{k_j-\lambda_\alpha+2iS\gamma}
{k_j-\lambda_\alpha-2iS\gamma}, \nonumber \\
&&\prod_{j=1}^N \frac{\lambda_\alpha-k_j-2iS\gamma}
{\lambda_\alpha-k_j+2iS\gamma} = \prod_{\beta,\beta\neq\alpha}^M
\frac{\lambda_\alpha-\lambda_\beta-2i\gamma}{\lambda_\alpha-
\lambda_\beta+2i\gamma},
\end{eqnarray}
where $\lambda_\alpha$ are the rapidities of the spin waves and 
$M$ is the total number of spin wave quanta. Here we dropped a 
multiplying factor $\exp[i\pi (2S+1)N]$ in the first of  Eqs.(11), 
because it is irrelevant in the thermodynamic limit. The first set 
of Eqs.~(11) is determined by the eigenvalue of the transfer matrix 
$tr_0 T_0(k_j)=\exp(-ik_jL)$, while the second set arises as the 
condition that ensures that the Bethe states are eigenstates of 
the Hamiltonian. The energy eigenvalues are given by
\begin{eqnarray}
E=\sum_{j=1}^N k_j^2,
\end{eqnarray} 
by acting the Hamiltonian on the asymptotic wave function.

\subsection{Model $II$}

For our second model the two-body $S$-matrix is given by
\begin{eqnarray}
&&S_{i,j}=-\frac{\Gamma(1+ik_{ij})}{\Gamma(1-ik_{ij})}
\frac{\Gamma(1+2S-{\hat l}_{ij}-ik_{ij})}
{\Gamma(1+2S-{\hat l}_{ij}+ik_{ij})}P_{ij}. 
\end{eqnarray}
With the same procedure, we obtain the BAE as
\begin{eqnarray}
&&e^{ik_jL}=\prod_{\alpha=1}^M \frac{k_j-\lambda_\alpha-2iS\gamma}
{k_j-\lambda_\alpha+2iS\gamma},\nonumber\\
&&\prod_{j=1}^N\frac{\lambda_\alpha-k_j-2iS\gamma}{\lambda_\alpha
-k_j+2iS\gamma}=\prod_{\beta,\beta\neq\alpha}^M\frac{\lambda_\alpha
-\lambda_\beta-2i\gamma}{\lambda_\alpha-\lambda_\beta+2i\gamma}.
\end{eqnarray}
The energy eigenvalues are still given by Eq.(12).

\section{Thermodynamics}

\subsection{model $I$}

Since $g_{ij}\geq 0$, the potential is repulsive and no charge 
bound state can exist. This can be seen clearly in the two-particle 
case. In fact, the two-particle wave function can be solved exactly 
in terms of two hypergeometric functions which has no bound state 
solution in the parameter region we considered. The solutions of 
the BAE, Eqs.(11), are therefore classified into real charge 
rapidities $k_j$ and strings of length $n$ of spin-rapidities.\cite{taka} 
In the thermodynamic limit $L\to\infty, N\to\infty, N/L\to n_e$, we 
denote with $\rho(k)$, $\sigma_n(\lambda)$ and $\rho_h(k)$, 
$\sigma_n^h(\lambda)$ the rapidity densities and the densities of 
their holes, respectively. By taking the thermodynamic limit of 
the BAE, we have
\begin{eqnarray}
\frac 1{2\pi}=\rho(k)+\rho_h(k)+\hat{B}'\rho(k)-\sum_n\hat{B}_{2S,n}
\sigma_n(k),\nonumber\\
\sigma_n^h(\lambda)+\sum_{m=1}^\infty \hat{A}_{m,n}\sigma_n(\lambda)
=\hat{B}_{2S,n}\rho(\lambda),
\end{eqnarray}
where the integral operators 
are given by ${\hat B}'=\sum_{m=1}^{2S}[2m]$, ${\hat B}_{2S,n}=
\sum_{l=1}^{min\{2S,n\}}[2S+n+1-2l]$, ${\hat A}_{m,n}=[m+n]+2[m+n-2]
+\cdots+2[|m-n|+2]+[|m-n|]$, and $[n]$ is the integral operator 
with the kernel 
\begin{eqnarray}
a_n(k) =\frac{ n\gamma }{\pi(k^2+n^2\gamma^2)}.
\end{eqnarray}
The free energy is expressed as
\begin{eqnarray}
&&F/L=\int(k^2-\mu-Sh)\rho(k)dk+\sum_n nh\int \sigma_n(\lambda)
d\lambda\nonumber\\
&&-T\int[(\rho(k)+\rho_h(k))\ln(\rho(k)+\rho_h(k))-\rho(k)\ln\rho(k)
-\rho_h(k)\ln\rho_h(k)]dk\\
&&-T\sum_n\int[(\sigma_n(\lambda)+\sigma_n^h(\lambda))
\ln(\sigma_n(\lambda)+\sigma_n^h(\lambda))-\sigma_n(\lambda)
\ln\sigma_n(\lambda)-\sigma_n^h(\lambda)\ln\sigma_n^h(\lambda)]
d\lambda. \nonumber
\end{eqnarray} 
where $\mu, T$ and $h$ are the chemical potential, the temperature 
and the magnetic field, respectively. It is convenient to introduce 
the quantities $\zeta(k)=\rho_h(k)/\rho(k)$ and $\eta_n(\lambda)
=\sigma_n^h(\lambda)/\sigma_n(\lambda)$. By minimizing the free 
energy with respect to the particle dendities, we obtain\cite{yang2}
\begin{eqnarray}
\ln\zeta=\frac{k^2-\mu-Sh}T+{\hat B}'\ln(1+\zeta^{-1}) 
-\sum_{n=1}^\infty {\hat B}_{2S,n}\ln(1+\eta_n^{-1}) , \\
\ln(1+\eta_n)=\frac{nh}T+\sum_{m=1}^{\infty}{\hat A}_{m,n}
\ln(1+\eta_m^{-1})-{\hat B}_{2S,n}\ln(1+\zeta^{-1}). \nonumber
\end{eqnarray}
Substituting the above equations into Eq.(17), we obtain
\begin{eqnarray}
F/L=-\frac T{2\pi}\int_{-\infty}^\infty\ln[1+\zeta^{-1}(k)]dk.
\end{eqnarray}
The thermodynamic BAE (18) can be further simplified. For 
convenience, we define $\hat{G}=[1]/([0]+[2])$. With the 
following operator identities
\begin{eqnarray}
\hat{A}_{m,n}-\hat{G}[\hat{A}_{m,n+1}+\hat{A}_{m,n-1}]&=&
\delta_{m,n},\nonumber\\
\hat{A}_{1,m}-\hat{G}\hat{A}_{2,m}&=&\delta_{1,m},\nonumber\\
\hat{B}_{m,n}-\hat{G}[\hat{B}_{m,n+1}+\hat{B}_{m,n-1}]&=&
\delta_{m,n}\hat{G},\\
\hat{B}_{1,m}-\hat{G}\hat{B}_{2,m}&=&\delta_{1,m}\hat{G},
\nonumber
\end{eqnarray}
we can rewrite the second equation of Eqs.(18) as
\begin{eqnarray}
\ln\eta_n = - \delta_{n,2S}{\hat G}\ln(1+\zeta^{-1}) 
+{\hat G}[\ln(1+\eta_{n+1})+\ln(1+\eta_{n-1})],
\end{eqnarray}
with the boundary conditions $\eta_0=0$, and $\lim_{n\to\infty}
\ln\eta_n/n=h/T\equiv 2x_0$. From Eqs.(20) obviously $\hat{G}
\hat{A}_{m,n}=\hat{B}_{m,n}$, so that from the second of Eqs.(18) 
we have
\begin{eqnarray}
\hat{G}\ln(1+\eta_{2S})=\frac{Sh}T+\sum_m\hat{B}_{m,2S}
\ln(1+\eta_m^{-1})-\hat{G}\hat{B}_{2S,2S}\ln(1+\zeta^{-1}).
\end{eqnarray}
Substituting the above equation into the first equation of 
Eqs.(18), we get
\begin{eqnarray}
\ln\zeta &=& \frac{k^2-\mu}T+{\hat B}'\ln(1+\zeta^{-1}) 
- {\hat G}[{\hat B}_{2S,2S}\ln(1+\zeta^{-1})+\ln(1+\eta_{2S})]
\nonumber \\
&=& \frac{k^2-\mu}T+{\hat B}''\ln(1+\zeta^{-1}) - {\hat G}
\ln(1+\eta_{2S}),  
\end{eqnarray}
where $\hat{B}''=\hat{G} \sum_{m=1}^{2S} [2m+1]$.

For $T \to 0$ and $h \to 0$ we obtain from Eq.(23) that $\zeta 
\to 0$ for $k<\sqrt\mu$ and $\zeta \to \infty$ for $k>\sqrt\mu$. 
This implies that the charges form a Fermi sea. In addition, in 
the same limit $\eta_{2S} \to 0$ while all other $\eta_n$ tend to 
some constant as indicated by Eq.(21). Therefore, the spin quanta 
form a Fermi sea of $2S$-strings in the ground state, as in the 
case of the BTM. The ground state properties are then given in 
terms of $\rho(k)$ and $\sigma_{2S}(\lambda)$ (which describe the 
two Fermi seas). The rapidities are densely distributed in the 
intervals $[-K,K]$ and $[-\Lambda,\Lambda]$, respectively, where 
in zero-field $\Lambda \to \infty$ for finite $\gamma$, and the 
integral equations are
\begin{eqnarray}
\rho(k)=\frac1{2\pi}-\int_{-K}^K B'(k-k')\rho(k')dk'
+\int_{-\Lambda}^\Lambda B_{2S,2S}(k-\lambda)\sigma_{2S}(\lambda)
d\lambda,\\
\int_{-\Lambda}^\Lambda A_{2S,2S}(\lambda-\lambda')
\sigma_{2S}(\lambda')d\lambda' = \int_{-K}^K B_{2S,2S}(\lambda-k)
\rho(k)dk , \nonumber
\end{eqnarray}
where $B'(k)$, $B_{m,n}(k)$ and $A_{m,n}(\lambda)$ are the kernels 
of ${\hat B}'$, ${\hat B}_{m,n}$ and ${\hat A}_{m,n}$, respectively. 
The elementary excitations are uniquely determined by the dressed 
energies $\epsilon_c(k)$ (of charges) and $\epsilon_s(\lambda)$ 
(of spins), which for $h \to 0$ satisfy
\begin{eqnarray}
\epsilon_c(k)=k^2-\mu-\int_{-K}^K B'(k-k')\epsilon_c(k')dk'
+\int_{-\Lambda}^\Lambda B_{2S,2S}(k-\lambda)\epsilon_s(\lambda)
d\lambda, \\
\int_{-\Lambda}^\Lambda A_{2S,2S}(\lambda-\lambda')
\epsilon_s(\lambda')d\lambda'=\int_{-K}^K B_{2S,2S}(\lambda-k)
\epsilon_c(k)dk , \nonumber
\end{eqnarray}
The group velocities of the two bands are $v_c = \epsilon_c'(K) 
/2\pi \rho(K)$ and $v_s = \epsilon_s'(\Lambda) /2\pi 
\sigma_{2S}(\Lambda)$, where the prime denotes derivative with 
respect to the rapidity. By integrating the second of Eqs. (25) 
for $h=0$ ($\Lambda=\infty$) we have $N=2M$, i.e. as expected the 
ground state is a spin singlet. At low $T$ the system then is a 
two-component Luttinger liquid with charge dynamics of central 
charge $1$ and spin dynamics with central charge $2S$. The latter 
is very similar to the BTM. 
 
In the Calogero-Sutherland limit, all the above quantities can be 
derived explicitely, since for $\gamma\to 0$, the integral operator 
$[m]$ becomes a $\delta$-function,
\begin{eqnarray}
\rho(k)=2\sigma_{2S}(k)=\frac{\theta(K-|k|)}{2\pi(S+1)},\nonumber\\
\epsilon_c(k)=2\epsilon_s(k)=\frac{k^2-K^2}{S+1},\\
v_c=v_s=2\pi n_e(S+1),\nonumber
\end{eqnarray}
where $\theta(x)$ is the step function and the Fermi momentum is 
$K=\pi n_e(S+1)$ with $n_e = N/L$. Without magnetic field the 
particles occupy each spin state equally, so that the distribution 
function $n_s(k)$ ($s=-S,\cdots, S$) is that of an exclusion statistics 
with the statistical weight $g=(S+1)(2S+1)$,
\begin{eqnarray}
n_s(k)=\frac{\theta(K-|k|)}{(S+1)(2S+1)}.
\end{eqnarray}

\subsection{Model $II$}

The solutions of the BAE for model $II$ are still given by real $k_j$ 
and $\lambda$-strings of length $n$. In the thermodynamic limit, the 
BAE, Eqs.(14), yield
\begin{eqnarray}
\frac1{2\pi}=\rho(k)+\rho_h(k)+\sum_{n=1}^\infty\hat{B}_{n,2S}
\sigma_n(k),\\
\sum_{m=1}^\infty\hat{A}_{m,n}\sigma_m(\lambda)+\sigma_n^h(\lambda)
=\hat{B}_{2S,n}\rho(\lambda).\nonumber
\end{eqnarray}
The free energy is still given by Eqs.(17) and (19). Using a similar 
procedure, i.e., by minimizing the the free energy functional with
respect to $\rho(k)$ and $\sigma_n(\lambda)$, we readily obtain 
the thermal BAE of model $II$
\begin{eqnarray}
\ln\zeta &=& \frac{k^2-\mu-Sh}T - \sum_{n=1}^\infty {\hat B}_{2S,n} 
\ln(1+\eta_n^{-1}) , \nonumber \\
\ln(1+\eta_n) &=& \frac{nh}T + \sum_{m=1}^{\infty} {\hat A}_{m,n}
\ln(1+\eta_m^{-1}) + {\hat B}_{2S,n} \ln(1+\zeta^{-1}), 
\end{eqnarray}
or equivalently
\begin{eqnarray}
\ln\zeta=(k^2-\mu)/T-{\hat G}\ln(1+\eta_{2S})
-{\hat G}{\hat B}_{2S,2S}\ln(1+\zeta^{-1}) , \\
\ln\eta_n=\delta_{n,2S}{\hat G}\ln(1+\zeta^{-1})
+{\hat G}\ln[(1+\eta_{n+1})(1+\eta_{n-1})]\nonumber,
\end{eqnarray}
with the same boundary conditions as in Eq.(21).  A strikingly 
different feature of this model is that when $T\to 0$, $h\to0$, all 
$\eta_n\to\infty$, which indicates that there are no flipped spins 
in the ground state. Hence, the ground state is ferromagnetic, since 
all the electrons align along a spin direction. The rapidity 
distributions in the ground state are then
\begin{eqnarray}
\rho(k)=\frac1{2\pi}\theta(K-|k|),\\
\sigma_n(\lambda)=0,\nonumber
\end{eqnarray}
where $K$ is the cutoff (integration limit or Fermi momentum) given
by
\begin{eqnarray}
K=\pi n_e.
\end{eqnarray}
The group velocity of the charge excitations at the Fermi surface is
\begin{eqnarray}
v_c=2\pi n_e.
\end{eqnarray}
The spin sector is no longer a Luttinger liquid, but has a quantum 
critical point. A single spin-wave excitation with momentum $p$, has 
an excitation energy $\epsilon_s(p)\sim p^2$ if $p$ is small. At low 
but finite $T$, the charge sector behaves almost like non-interacting 
spinless fermions, while for the spin sector we have a ferromagnetic
Babujian-Takhatajan spin liquid, so that asymptotically as $T \to 0$ 
the specific heat and  the spin susceptibility are \cite{schl}
\begin{eqnarray}
C\sim T^{\frac12},{~~}\chi\sim T^{-2}\ln^{-1}T .
\end{eqnarray}
The low-$T$ specific heat can also be inferred from the spin-wave 
dispersion relation. 

In the Calogero-Sutherland limit $\gamma \to 0$, bothm, the charge 
and spin sectors, obey ideal fractional statistics. The thermal BAE 
can then be solved analytically, since in this limit the integral 
equations reduce to algebraic ones. Solving for $\eta_n$ separately 
for $n\leq 2S$ and $n\geq 2S$ we obtain
\begin{eqnarray}
\eta_n&=&\frac{\sin^2[(n+1)\alpha]}{\sin^2\alpha}-1, \ \ \ n\leq 2S, 
\nonumber \\
\eta_n&=&\frac{\sinh^2[(n-2S)x_0+\beta]}{\sinh^2x_0}-1, \ \ \ 
n\geq 2S,
\end{eqnarray}
where $\alpha$ and $\beta$ are functions of $\zeta(k)$ determined by
\begin{eqnarray}
{\sin[(2S+1)\alpha]}{\sinh x_0}={\sinh\beta}{\sin\alpha},\nonumber\\
\sin(2S\alpha)\sqrt{1+\zeta^{-1}}=\frac{\sin\alpha}{\sinh x_0}
|\sinh(\beta-x_0)|.
\end{eqnarray}
The solution of Eqs.~(35) and (36) yields $\eta_n$ as a function of 
$\zeta$. Substituting the result into the first of Eqs.(30) we obtain 
a relation for $\zeta(k)$, which uniquely determines the free energy. 
The distribution function for the charges is 
\begin{eqnarray}
n(k)=-\frac{\partial}{\partial \mu}\ln[1+\zeta^{-1}(k)].
\end{eqnarray}
A careful calculation also shows that the logarithm in Eq.(34) 
disappears, i.e., $\chi\sim T^{-2}$. This is due to the free nature 
of the particles in the limit $\gamma\to 0$.

\section{Concluding remarks}

In conclusion, we constructed two integrable models with long-range 
spin couplings. In model $I$, the coupling between spins is of the 
antiferromagnetic exchange form, while in model $II$, the spin 
coupling is ferromagnetic of the Babujian-Takhatajan type. The BAE 
for the spin sector are the same as for the BTM (here the $k_j$ 
introduce a lattice disorder). As we have learned from ordinary 
integrable models, almost all lattice models have a continuum 
counterpart in the sense that their BAE share a common structure. 
Examples are the spin-1/2 isotropic Heisenberg chain\cite{bethe} 
with the $\delta$-potential boson gas model,\cite{lieb} the Hubbard 
model\cite{li-wu} with the $\delta$-potential Fermi gas model, 
\cite{yang} and the $SU(N)$-invariant $t-J$ model\cite{suth2} 
with the $SU(N)$-invariant $\delta$-potential Fermi gas model. 
\cite{suth3} However, no continuum counterpart to the BTM with 
contact interaction was known so far.\cite{andr} The present work 
fills this gap in the integrable model family. 

Finally, we conjecture that the following lattice model 
\begin{eqnarray}
H=\sinh^2\gamma\sum_{m<n}^N\frac{ Q({\vec S}_m\cdot{\vec S}_n)}{
\sinh^2[\gamma(m-n)]}, \label{newH}
\end{eqnarray}
is integrable for the following reasons: (a) With the proper freezing 
process (eliminating the charge dynamics), the continuum models may 
be reduced to Eq.(\ref{newH}), since the BAE have the same form as 
for the BTM.\cite{note} (b) For $\gamma\to\infty$, Eq.(\ref{newH}) 
flows to the BTM, which has been demonstrated to be integrable.

We acknowledge the support by the National Science Foundation 
and the Department of Energy under grants No. DMR98-01751 and
No. DE-FG02-98ER45797. Y. Wang is also supported by the 
National Science Foundation of China.


\begin{references}
\bibitem{note}The freezing procedure in our case is not trivial 
because of the absence of a free parameter as in the $SU(N)$ case. 
With such a free parameter both models may become special limits of 
an unifying Hamiltonian.
\end{references}
\end{document}